# Cox-nnet v2.0: improved neural-network based survival prediction extended to large-scale EMR dataset

Di Wang[1], Kevin He[1], Lana X Garmire[2*]

[1]Department of Biostatistics, University of Michigan, Ann Arbor, MI

[2]Department of Computational Medicine and Bioinformatics, University of Michigan, Ann Arbor, MI

[*]Corresponding author: lgarmire@med.umich.edu

## Abstract

**Summary:** Cox-nnet is a neural-network based prognosis prediction method, originally applied to genomics data. Here we propose the version 2 of Cox-nnet, with significant improvement on efficiency and interpretability, making it suitable to predict prognosis based on large-scale electronic medical records (EMR) datasets. We also add permutation-based feature importance scores and the direction of feature coefficients. Applying on an EMR dataset of OPTN kidney transplantation, Cox-nnet v2.0 reduces the training time of Cox-nnet up to 32 folds (n=10,000) and achieves better prediction accuracy than Cox-PH ($p<0.05$).

**Availability and implementation:** Cox-nnet v2.0 is freely available to the public at

https://github.com/lanagarmire/Cox-nnet-v2.0

## Introduction

Large-scale Electronic medical records (EMR) are informative and easily accessible data sources frequently used for patients survival prediction. Prediction models built on EMR data tend to have better performance than those using administrative data (Mahmoudi *et al.*, 2020). It is also found that machine learning based models outperformed conventional models, such as Cox-Proportional Hazard (Cox-PH) model (Cox, 1972), Random Survival Forests (RSF) model (Ishwaran *et al.*, 2008) and elastic net regression (Fan *et al.*, 2010) on the prediction of coronary artery disease mortality using EMR data (Steele *et al.*, 2018). Although it is challenging to develop prediction models driven by EMR data, the large sample size and clinical features in EMR data provide valuable information in survival prediction (Goldstein *et al.*, 2017).

We previously proposed Cox-nnet (Ching *et al.*, 2018), a deep learning based neural network prognosis prediction model, which achieved comparable or better performance than Cox-PH on high-throughput omics data. We recently applied Cox-nnet to histopathology imaging data with pre-extracted features, and demonstrated its advantage in combining gene expression data and image data for survival prediction (Zhan *et al.*) . However, it remains to be tested if Cox-nnet is suitable to predict survival in large-scale EMR data, where the patient size is usually magnitudes larger than genomics data. Towards this, we propose Cox-nnet v2.0, which significantly improves computational speed, with enhanced interpretability. Additionally, Cox-nnet v2.0 also achieves better prediction accuracy than Cox-PH.

## Methods

***Cox-nnet method improvement***: The original Cox-nnet is a neural-network based extension to Cox-PH method, using the log partial likelihood as its loss function. In Cox-nnet v2.0, we have made the following improvements:

(1) Speed-up in calculating log partial likelihood loss function. The log partial likelihood is calculated by:

$$pl(\boldsymbol{\beta}) = \sum_{C_i=1} (\boldsymbol{\theta}_i - \log \sum_{t_i \leqslant t_j} exp(\boldsymbol{\theta}_i))$$

Where $\boldsymbol{\theta}_i$ is the linear predictor of patient $i$ and $C_i$ is defined by $C_i = I$(patient $i$ is not censored). To avoid nested summation in Theano, the previous version of Cox-nnet calculates the log partial likelihood by matrix multiplication:

$$pl(\boldsymbol{\beta}) = \{\boldsymbol{\theta} - log(R * exp(\boldsymbol{\theta}\ ))\}^T C$$

Where $C$ and $\boldsymbol{\theta}$ are vectors of $C_i$ and $\boldsymbol{\theta}_i$ respectively. $R$ is a $n$ by $n$ at risk set indicator matrix, and each entry $R_{ij}$ is defined by:

$$R_{ij} = I(t_i \leq t_j)$$

Where $n$ is the sample size of the input data, and $t_i$ and $t_j$ are the event time of patient $i$ and $j$, respectively. This implementation is memory intensive and time consuming when dealing with large sample sizes.

In the new version, instead of pairwise comparison we sorted the observations by event time. Then by definition of the at risk set, $R$ is converted to an upper triangular matrix filled with 1. Intuitively, $R * exp(\boldsymbol{\theta})$ can be calculated using cumulative summation that no longer requires storing $R$ matrix and nested summation (double loops).

(2) Adding permutation based feature importance scores. Previously the variable importance score of Cox-nnet is calculated by pseudo drop-out, which replaced the variable with its mean. The drawback is that it is hard to interpret categorical variables. Here we introduce a more general feature evaluation method using permutation importance score (Breiman, 2001). The main idea is to measure the model error increase after shuffling the feature's values, since the permutation breaks the relationship between the feature and the outcome. We implement the algorithm proposed in Fisher *et al.* (2019).

(3) Adding the directionality of the feature coefficient. Similar to estimating the sign of $\beta$ for Cox-PH, we develop a framework which approximates the direction of feature coefficients in Cox-nnet. The linear predictor in Cox-nnet is:

$$\boldsymbol{\theta}_i = G(WX_i + b)\beta$$

Where $G$ is the activation function, $W$ is the coefficient weight matrix between input and hidden layer, and $b$ is the bias term. Suppose each column $X_k^*$ in $X^*$ is defined by:

$$X_k^* = (X_k - 1) \times I(X_k \text{ is continuous variable}) + 0 \times I(X_k \text{ is categorical variable})$$

Similar to the interpretation of $\beta$ in Cox-PH, the direction of each feature coefficient in Cox-nnet is approximated by the sign of

$$\frac{1}{n}\sum_{i=1}^{n} \Delta\boldsymbol{\theta}_{ik} = \frac{1}{n}\sum_{i=1}^{n} (\boldsymbol{\theta}_i - \boldsymbol{\theta}_{ik}^{**}) = \frac{1}{n}\sum_{i=1}^{n} \{G(WX_i + b)\beta - G(WX_{ik}^{**} + b)\beta\}.$$

Where $X_{ik}^{**}$ is defined by $X_{ik}^{**} = (X_{ik}^*, X_{i(-k)})$. Intuitively, the risk increases if the sign of

$\frac{1}{n}\sum_{i=1}^{n} \Delta\boldsymbol{\theta}_{ik}$ is positive.

(4) Adding additional optimization algorithms and activation functions for parameter tuning. We add Adam (Kingma and Ba, 2014) optimizer as an alternative optimization strategy, which further accelerates the training process. We also use the Scaled Exponential Linear Unit (SELU) activation function (Klambauer *et al.*, 2017) in the Cox-nnet v2.0.

***Evaluation Metrics***: As in Cox-nnet v1.0, we evaluate the prediction accuracy by C-IPCW (Uno *et al.*, 2011), which is the C-index weighted by inverse censoring probability.

***Dataset***: The EMR data used for the study is kidney transplant data obtained from the U.S. Organ Procurement and Transplantation Network (OPTN) (https://optn.transplant.hrsa.gov/data/). A total of 80,019 patients which includes all patients with ages greater than 18 who received transplant between January 2005 and January 2013 with deceased donor type were used in the analysis. A total of 117 clinical variables describing up-to transplant characteristics are used in the analysis.

## Results

The structure of Cox-nnet v2.0 is shown in Fig. 1A. The newly updated functionalities are highlighted. We randomly split the kidney transplant EMR data into training (80%) and testing (20%) sets, and used C-IPCW to evaluate on the hold-out testing set. We repeated this process 10 times to access the overall prediction performance. Cox-nnet v2.0 is not sensitive to the sample size and dramatically reduces the training time, compared to Cox-nnet v1.0 where the computing time increases polynomially with the sample size (Fig. 1B). Cox-nnet v2.0 also achieves significantly better C-IPCW than Cox-PH (Fig. 1C), without any drop of C-IPCW compared to Cox-nnet v1.0. We performed feature evaluation by calculating the feature importance scores using the new permutation method, where the values are close to those by the previous pseudo drop-out method. With the directionality (+/- signs) of the feature coefficients, our feature evaluation results are more interpretable: a positive (+) sign indicates increased risk of graft failure, whereas a negative (-) sign means decreased risk of graft failure. As additional confirmation, the pattern of important scores matches well with that of coefficients obtained from Cox-PH (Fig. 1D).

In summary, Cox-nnet v2.0 significantly accelerates the training process of Cox-nnet without loss in the prediction accuracy. In addition, it also enables better interpretation for all features in the model. Cox-nnet v2.0 is the new version suitable for prognosis prediction in large-scale EMR dataset.

## Author's Contribution

LXG conceived the project, DW conducted model improvement and data analysis, KH provided the dataset and helped with the analysis. DW and LXG wrote the manuscript with the help of KH. All authors read, revised and approved the manuscript.

## Declaration of Conflict of Interest

The authors declare no conflict of interest.

## Acknowledgement

LXG is support by grants K01ES025434 awarded by NIEHS through funds provided by the trans-NIH Big Data to Knowledge (BD2K) initiative (www.bd2k.nih.gov), R01 LM012373 and R01 LM012907 awarded by NLM, and R01 HD084633 awarded by NICHD. KH is supported by funds through The University of Michigan Office of Research (UMOR).

**Figure 1. Overview of Cox-nnet v2.0 and its performance improvement.**
(A) The modules in Cox-nnet. The names of new modules and functions are in bold with highlight background. The other modules are inherited from Cox-nnet v1.0 (B) Training time comparison among Cox-nnet v2.0 (red), Cox-nnet v1.0 (green) and Cox-PH (purple), varying from sample size of 1,000 to 10,000. (C) Prediction accuracy measured by C-IPCW on the EMR dataset (n=80,019), over 10 repetitions. *: P<0.05 (1-tail t-test)  (D). Heatmap to compare feature importance scores in different methods. From top to bottom row: pseudo-drop out (Cox-nnet v1.0), permutation importance score (Cox-nnet v2.0), permutation importance score times direction of feature coefficient (Cox-nnet v2.0), and scaled z-score (Cox-PH).

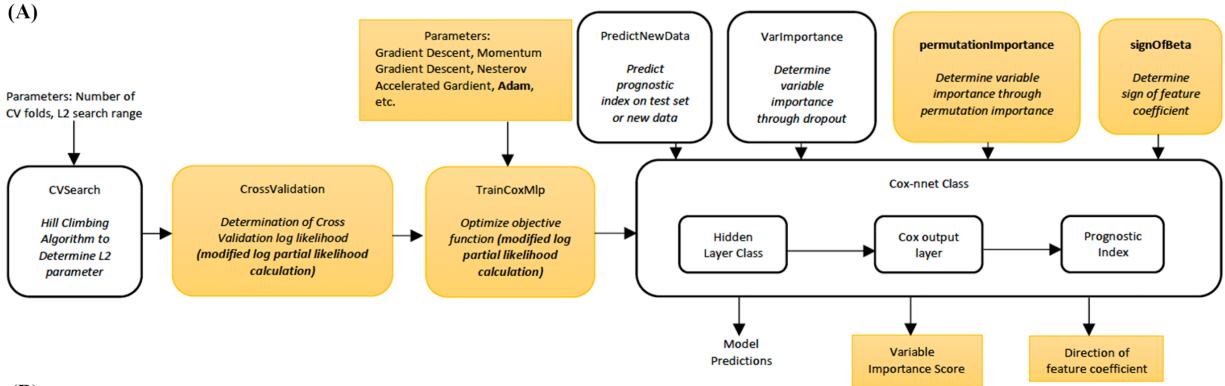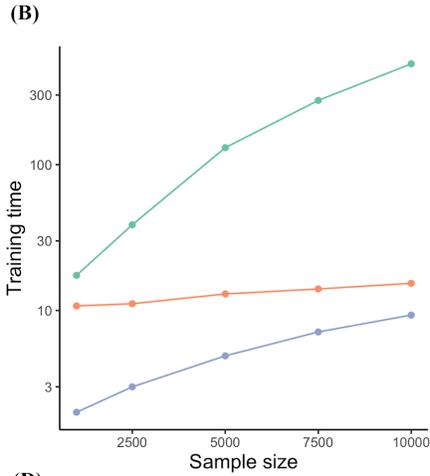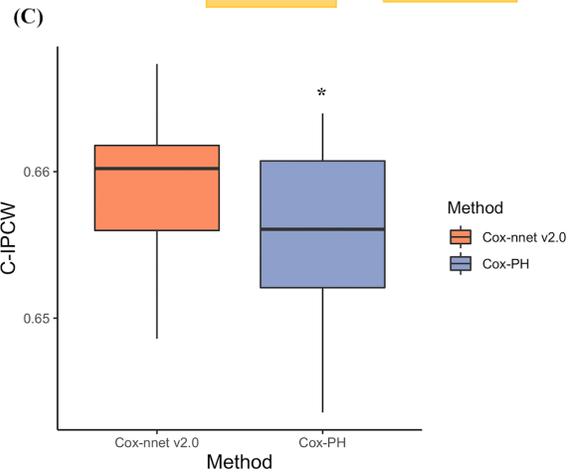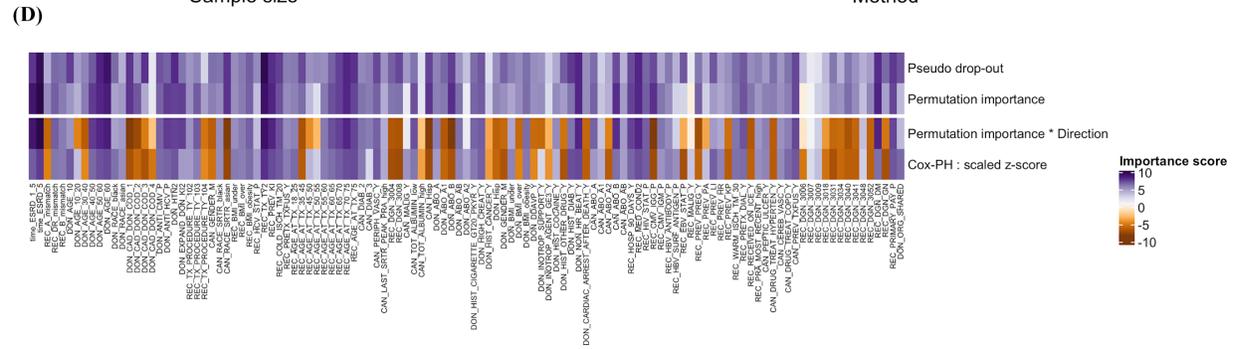